\documentclass[12pt,preprint]{aastex}
\def\a4{\hsize 17.0cm \vsize 25.cm}
\newcommand{\der}[2]  { \frac{{\rm d}#1}{{\rm d}#2} }

\shorttitle{superwinds \& further star formation}
\shortauthors{Silich et al.}

\begin{document}

\title{Superstellar clusters and their impact on their host galaxies.}

\author{
Guillermo Tenorio-Tagle, Sergiy Silich }
\affil{Instituto Nacional de Astrof\'\i sica Optica y
Electr\'onica, AP 51, 72000 Puebla, M\'exico; gtt@inaoep.mx}

\author{Casiana Mu\~noz-Tu\~n\'on}
\affil{Instituto de Astrof\'{\i}sica de Canarias, E 38200 La
Laguna, Tenerife, Spain; cmt@ll.iac.es}

%\altaffiltext{1}{Permanent address: Main Astronomical Observatory National 
%                 Academy of Sciences of Ukraine, 03680 Kiyv-127, Golosiiv, 
%                 Ukraine}

\begin{abstract}
We review the properties of young superstellar clusters and the impact that their evolution has
in their host galaxies. In particular we look at the two different star-forming feedback modes: positive and negative
feedback. The development of strong isotropic winds 
emanating from massive clusters, capable of disrupting the remains of the 
parental cloud as well as causing the large-scale restructuring of the surrounding ISM, has usually been taken as
a negative feedback agent. Here we show the impact that radiative cooling has on the resultant outflows
and then, as an extreme example, we infer from the observations of M82 the detailed inner structure of
supergalactic winds and define through numerical simulations the ingredients required to match such structures.
We also show how when radiative cooling becomes significant  within the 
star cluster volume itself ($\sim$  30$\%$ of the deposited energy),  
the force of gravity takes over and drives in situ all the matter deposited by winds and 
supernovae into several generations of star formation. A situation in which 
the mass deposition rate
$\dot M_{SC}$, instead of causing a wind as in the adiabatic solution, turns into
a positive feedback star-forming mode equal to the star formation rate. 
\end{abstract}

\section{The properties of superstellar clusters}

The discovery by HST  of a large population of unusually compact young superstellar clusters (SSCs) within starburst galaxies
(see review by  Ho 1997; Johnson et al. 2001; Colina et al. 2002; Larsen \& Richtler 2000, 
Kobulnicky \& Johnson 1999 and the proceedings edited by 
Lamers et al. 2004), 
has led  us to infer  the unit  of massive or violent star formation.
SSCs with masses in the range of a few $\times 10^5$ M$_\odot$ to up to 6 $\times 10^7$ M$_\odot$ 
(see Walcher et al. 2004; Pasquali et al.
2004) within a small volume of radius 3 to 10 pc, are indeed some of the most energetic entities 
found now in a large variety of  galaxies. Note also that collections of them have now been found within a 
single starburst galaxy, as in M82 and the antennae galaxy (see Melo et al 2005; Whitmore et al. 1999). 

The potential impact that these new 
units of star formation may have on the ISM of their host galaxies has been inferred from cluster synthesis models
(see e.g. Cervi\~no \& Mas-Hesse 1994,  Leitherer et al 1999).
A coeval cluster with 10$^6$ M$_\odot$ in stars, an initial mass function (IMF) similar  to that proposed by Salpeter and an
upper and lower mass range for the coeval event between 100 M$_\odot$ and 1 M$_\odot$, leads initially to several tens  
of thousands of O stars. These however begin to disappear rather quickly (after t $\sim$ 3 Myr) 
as they complete their evolution and explode as
supernovae (SNe). The cluster evolution is so rapid that after 10 Myr there are no O stars left within the cluster.   
All massive stars undergo strong stellar winds and all of them with a mass larger than 8 M$_\odot$ will end 
their evolution exploding as supernova. Thus, one is to expect from our
hypothetical cluster several tens of thousands of SNe over a time span of some 40 Myr. 
During the supernova phase a 10$^6$ M$_\odot$ stellar cluster will 
produce an almost constant energy input rate of the order of 10$^{40}$ erg s$^{-1}$. On the other hand, the ionizing luminosity 
emanating from the  cluster 
would reach a constant value of 10$^{53}$  photons s$^{-1}$ during the first 3.5 Myr of evolution to then 
drastically drop (as t$^{-5}$) as the most massive members of the association explode as supernova. The rapid drop in the 
ionizing photon flux implies that after 10 Myr of evolution, the $UV$ photon output would have fallen by more than two
orders of magnitude from its initial value and the HII region that they may 
have originally produced would have drastically reduced its dimensions.
Thus the HII region lifetime is restricted to the first 10 Myr of the evolution and is much shorter than the 
supernova phase. It is important to realize that only 10$\%$ of the stellar mass goes into stars with a mass larger than 
10 M$_\odot$, however, it is this   10$\%$ the one that causes all the energetics from the starburst.
Being massive, although smaller in number, massive stars  also reinsert into the 
ISM, through their winds and SN explosions, almost 40$\%$ of the starburst total original mass. And thus from a starburst with 
an initial mass of 10$^6$ M$_\odot$ one has to expect a total of almost 4 $\times 10^5$ M$_\odot$ violently injected back into the ISM,
during the 4 $\times 10^7$ years that the SN phase lasts. From these, almost 
40,000 M$_\odot$ will be in oxygen ions and less than 1000 M$_\odot$ in iron
(see Silich et al. 2001).
 
One of the features of the stellar synthesis models regarding the energetics of coeval star clusters is that they fortunately 
scale linearly with the 
mass of the starburst. It is therefore simple to derive the properties of starbursts of different masses, for as long 
as they present the IMF, metallicity and stars in the same mass range considered by the models.

When dealing with the outflows generated by star clusters, another important intrinsic property 
is the metallicity of their ejected matter. This is a  strongly varying function of time, bound 
by the yields from massive stars and their evolution time. Thus, once the cluster IMF and the stellar mass limits are 
defined, the resultant metallicity of the ejected material is an invariant curve, independent of the cluster mass. 
Here we consider  coeval clusters
with a Salpeter IMF, and stars between 100 M$_\odot$ and 1 M$_\odot$, as well as
the evolutionary tracks with rotation of Maynet \& Maeder (2002)  and an instantaneous
mixing of the recently processed metals with the stellar envelopes of the progenitors
(see Silich et al. 2001 and Tenorio-Tagle et al. 2003, for an explicit description of the calculations). 
This leads to metallicity values (using oxygen as tracer)
that rapidly reach 14 Z$_\odot$ (see Figure 1), and although steadily decaying afterwards,
the metallicity remains above solar values for a good deal of the
evolution (for more than 20 Myr), to then fall to the original metallicity of the
parental cloud, once the oxygen yield has been delivered. 
One of the main effects of an enhanced  
metallicity of the ejecta is to boost its radiative cooling  
and in such a case, massive clusters 
may inevitably enter into the strong cooling regime,  to then find
their stationary superwinds totally inhibited (see section 4). 

%\newpage

%---------------------------------------------------------------
\begin{figure}[htbp]
\plotone{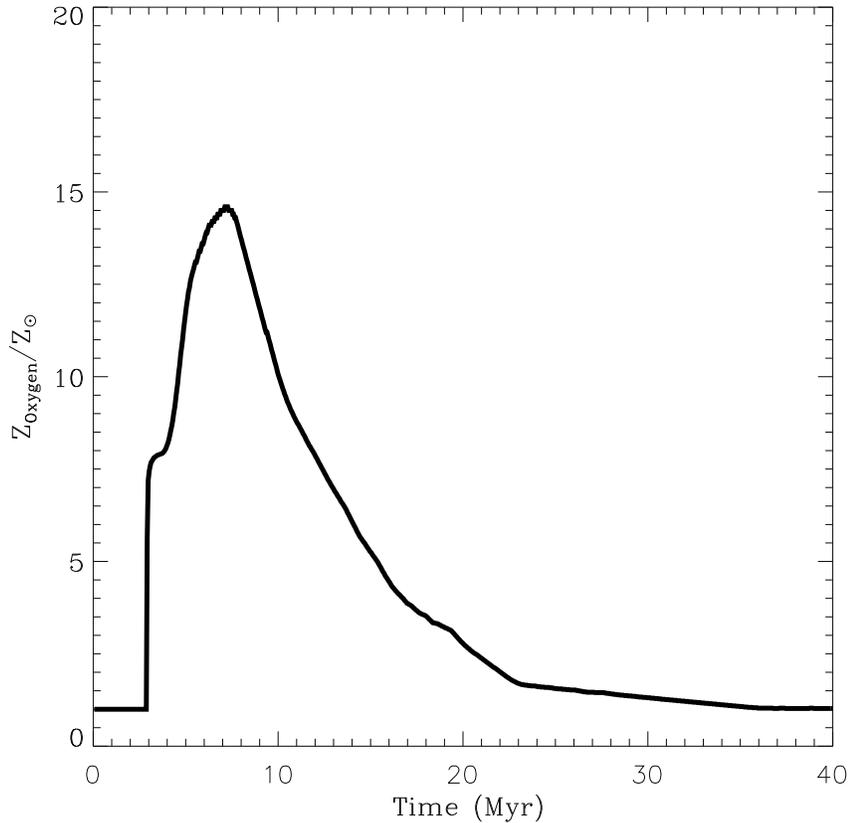}
\caption{The metallicity of the matter ejected by coeval bursts of
star formation. The metallicity (in solar units) of the matter
reinserted into the ISM by SCs is plotted as a function of the
evolution time. The curve is derived from the  metal yields of 
Meynet \& Maeder (2002) stellar evolution models with stellar 
rotation, using oxygen as a tracer. The estimate assumes a Salpeter 
IMF and 100 M$_\odot$ and 1 M$_\odot$ upper and lower mass limits.  
The model also assumes an instantaneous mixing between the newly 
processed metals and the envelopes of the progenitors. Note that under 
these assumptions the curve is independent of the cluster mass.}
\label{fig1}
\end{figure}
%---------------------------------------------------------------

%\newpage

\section{Feedback from superstellar clusters}

The close spacing between sources within a super-star cluster warrants a very efficient thermalization 
of all their winds and supernova explosions, leading to the high central overpressure that is to drive both a 
superbubble and in some cases a supergalactic wind (SGW). The outflow from the star cluster surface 
is fully defined by three quantities: The mass and mechanical energy deposition rates
(hereafter $\dot M_{SC}$ and $\dot E_{SC}$) and the radius that encompasses the newly born sources ($R_{SC}$).

In the adiabatic case, the total mass and energy deposition rates define the  temperature
and thus the sound speed $c_{SC}$ at the cluster surface.

%---------------------------------------------------------------------- 
\begin{equation}
      \label{eq.01} 
T_{SC} = \frac{0.3 \mu}{k} \frac{\dot E_{SC}}{{\dot M}_{SC}} , 
\end{equation} 
%----------------------------------------------------------------------

\noindent 
where $\mu$ is the mean mass per particle and $k$ the 
Boltzmann constant. On the other hand, the density of matter streaming out of $R_{SC}$
is:

%---------------------------------------------------------------------- 
\begin{equation}
\rho = \frac{{\dot M_{SC}}}{4 \pi R_{SC}^2 c_{SC}} ,
\end{equation} 
%----------------------------------------------------------------------

Thus at $R_{SC}$  (see Chevalier \& Clegg 1985; hereafter CC85), the ratio of thermal and kinetic energy flux to the total flux is

%---------------------------------------------------------------
\begin{equation}
      \label{eq.02} 
F_{th}/F_{tot} = \frac{\frac{1}{\gamma - 1}\frac{P}{\rho}}
                 {\frac{u^2}{2} + \frac{\gamma}{\gamma - 1}
                 \frac{P}{\rho}} = \frac{9}{20}
\end{equation} 
\begin{equation}
      \label{eq.03}   
F_{k}/F_{tot} = \frac{u^2/2}{\frac{u^2}{2} + \frac{\gamma}{\gamma - 1}
                 \frac{P}{\rho}} = \frac{1}{4} .
\end{equation} 
%-------------------------------------------------------------
There is however a rapid evolution as matter streams away from the
central starburst. After crossing $r = R_{SC}$ the gas is immediately 
accelerated across the steep pressure gradients and rapidly reaches 
its terminal velocity ($V_\infty \sim 2 c_{SC}$). This is due to a fast 
conversion of thermal energy, into kinetic energy of the resultant 
wind.

In a recent communication (Silich et al. 2003, 2004), we have revised the properties of SSCs by solving the flow equations 
without  the assumption of an adiabatic flow made by Chevalier \& Clegg (1985). In this case,
the steady-state solution results from solving 

%---------------------------------------------------------------
\begin{eqnarray}
      \label{eq.1a}
      & & \hspace{-0.5cm}
\frac{1}{r^2} \der{}{r}\left(\rho u r^2\right) = q_m ,
      \\[0.2cm]
      \label{eq.1b}
      & & \hspace{-0.5cm}
\rho u \der{u}{r} = - \der{P}{r} - q_m u
      \\[0.2cm]
     \label{eq.1c}
      & & \hspace{-0.5cm}
\frac{1}{r^2} \der{}{r}{\left[\rho u r^2 \left(\frac{u^2}{2} +
\frac{\gamma}{\gamma - 1} \frac{P}{\rho}\right)\right]} = q_e - Q,
\end{eqnarray}
%-------------------------------------------------------------
where $q_e$ and $q_m$ are the energy and mass deposition rates per unit volume
($q_e = q_m$ = 0 if $r$ exceeds $R_{SC}$), $Q$ is
the cooling rate  ($Q = n^2 \Lambda$) where $n$ is the wind number 
density and $\Lambda$ is the cooling function (a function of $T$ and metallicity). 
Central values of density and temperature are found by iteration, with the restriction that the flow velocity
ought to increase from 0 km s$^{-1}$ at the cluster center to the sound speed when it reaches the cluster surface.
If this happens then the stationary solution is fully warranted. A 
solution in which  what is put in ($\dot M_{SC}$) 
is at all times equal to the outflow from the star cluster surface 
(4 $\pi R_{SC}^2 \rho_{SC} c_{SC}$).
Radiative cooling leaves almost unaffected the wind density and
velocity distributions but its temperature may for energetic clusters
fall rapidly to $T \sim 10^4$ K, close to the SSC surface. Figure 2
compares the adiabatic (dotted lines) with the radiative solution
(solid lines) for a cluster  with a 
$\dot E_{SC}$ = 3 $\times 10^{41}$ erg s$^{-1}$, 
in which the wind temperature instead of steadily falling as $r^{-4/3}$, it 
rapidly falls to $T \sim 10^4$ K, reducing 
drastically the size of the resultant X-ray emitting volume.  
The energetic winds lead to a total restructuring of  the surrounding ISM,
and in some extreme cases their energetics may even reach the
outskirts of a galaxy causing a galactic wind. 

%\newpage

\begin{figure}[htbp]
\vspace{21.0cm}
\includegraphics{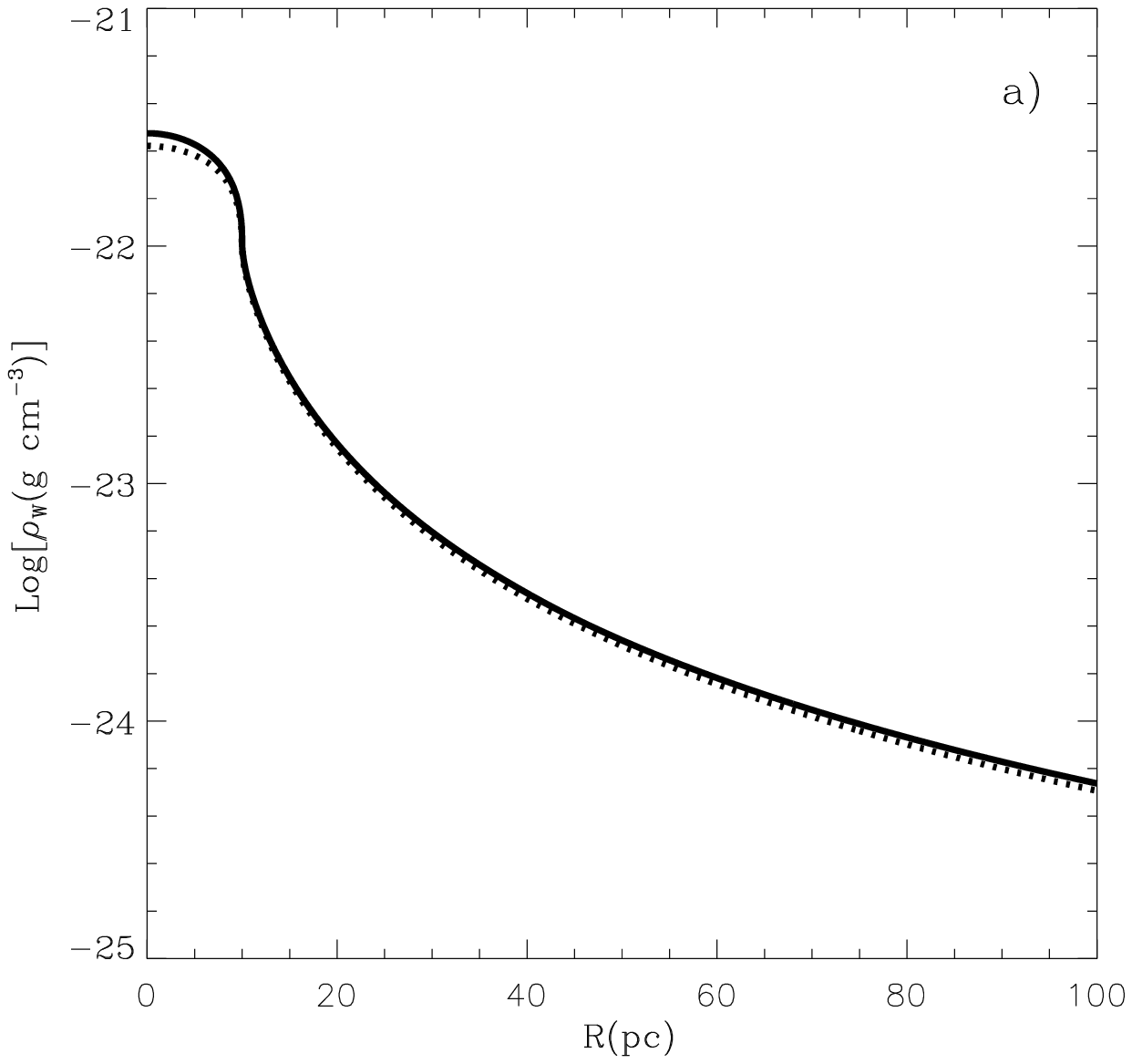}
\includegraphics{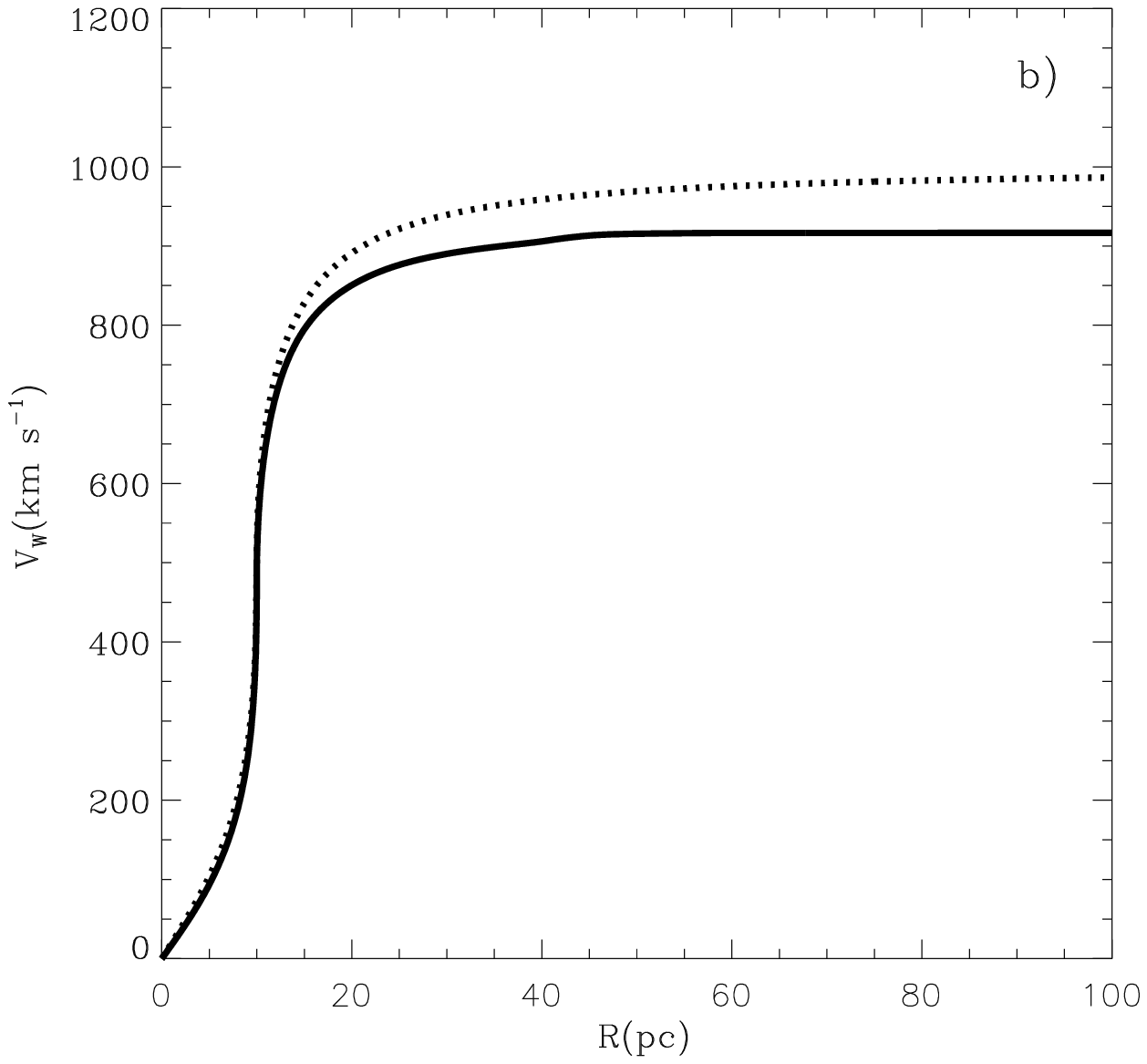}
\includegraphics{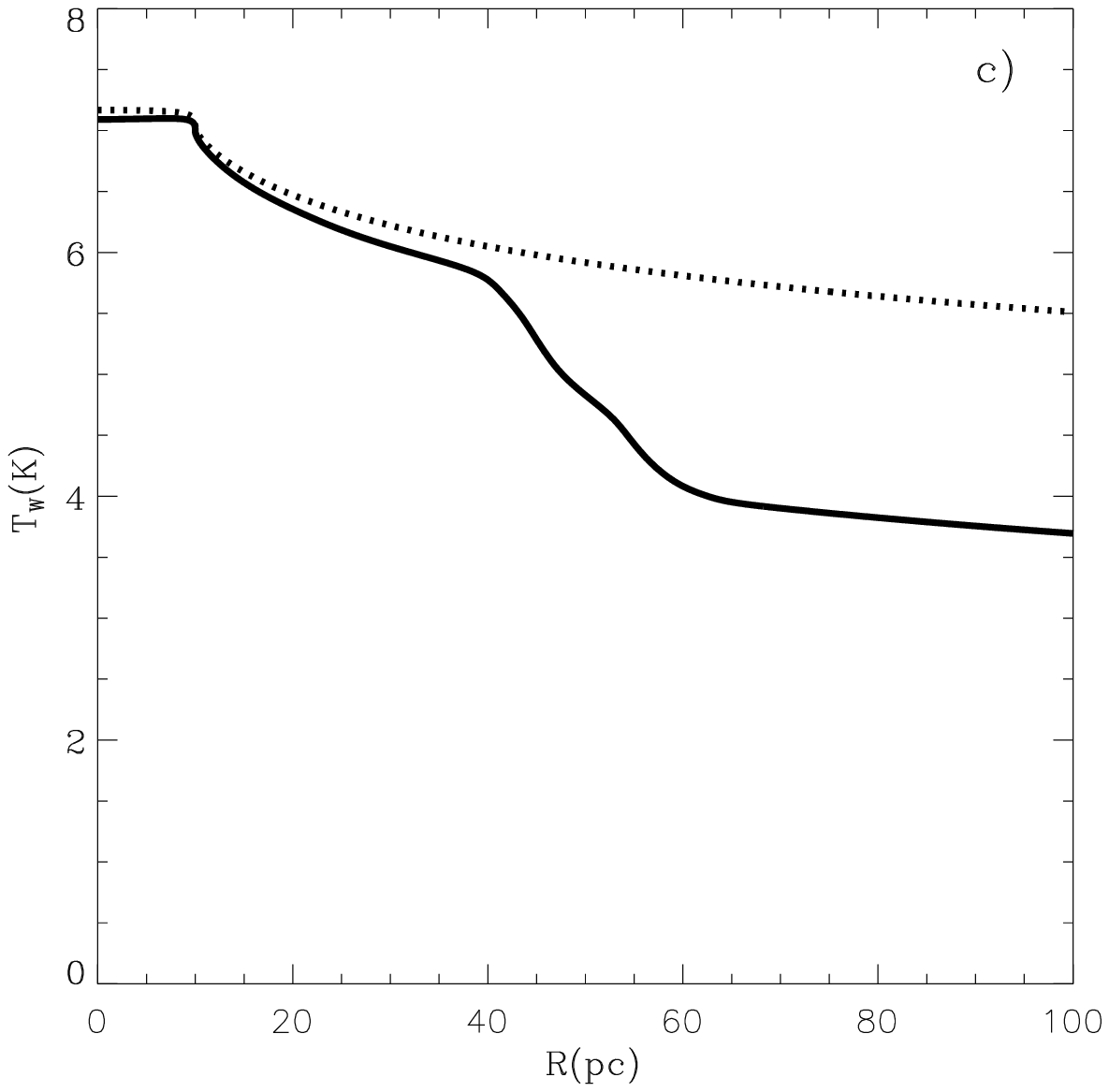}
%\plotone{fig2a.eps}
%\plotone{fig2b.eps}
%\plotone{fig2c.eps}
\caption{Adiabatic vs radiative superwinds. Density, velocity and 
temperature distributions of the wind produced by a 10 pc radius SSC 
that deposits 3 $\times 10^{41}$ erg s$^{-1}$ under the adiabatic 
(dotted lines) and radiative (solid lines) assumptions.}
\label{fig2}
\end{figure}
%---------------------------------------------------------------

%\newpage

We have also shown (see below section 4) that in the $\dot E_{SC}$ vs star cluster radius diagram, there is a threshold limit that 
massive and energetic compact clusters 
may cross to find out that radiative
cooling inhibits their  stationary outflow condition (Silich et al., 2004) and then the 
matter ejected by the stellar sources, unable to escape, is to accumulate within the  star 
cluster volume (see section 4).

\section{Negative feedback and the physics of supergalactic winds}

In all cases below the threshold line, which may be considered quasi-adiabatic or strongly radiative,
the energy dumped by the central starburst, is to cause a major impact on the surrounding gas. 
The supersonic stream leads immediately 
to a leading shock able to heat, accelerate and sweep all the overtaken material into a fast expanding shell. 
In this way, as the whole structure grows, the density, temperature and thermal pressure of the wind 
drops as $r^{-2}$, $r^{-4/3}$ and  $r^{-10/3}$, respectively (CC85).

Note however that such a flow is exposed to the appearance of reverse 
shocks whenever it meets an obstacle cloud or when its thermal pressure   
becomes lower than that of the surrounding gas, as it is the case in strongly radiative winds and within
superbubbles. There, the high pressure acquired by the swept up ISM 
becomes larger than that of the freely expanding ejecta (the free wind region; FWR), 
where $\rho$, $T$ and $P$ are rapidly falling.
The situation leads to the development of a reverse shock and with it to the thermalization of the wind
kinetic energy, reducing the size of the free-wind region. Thus for the FWR to extend 
up to large distances away from the host galaxy, the shocks  would have had
to evolve and displace all the ISM, leading to a free path into the 
intergalactic medium through which the free wind may flow as  a supergalactic wind. 
The energy required to achieve such a task, depends 
strongly on the ISM density distribution. As shown by Silich \& 
Tenorio-Tagle (2001) the energy required to burst into the
inter-galactic medium out of a fast rotating flattened galaxy is
orders of magnitude smaller than that required to exceed the
dimensions of a slow rotating and more spherical density distribution (see Figure 3).  

%\newpage

\begin{figure}[htbp]
\plotone{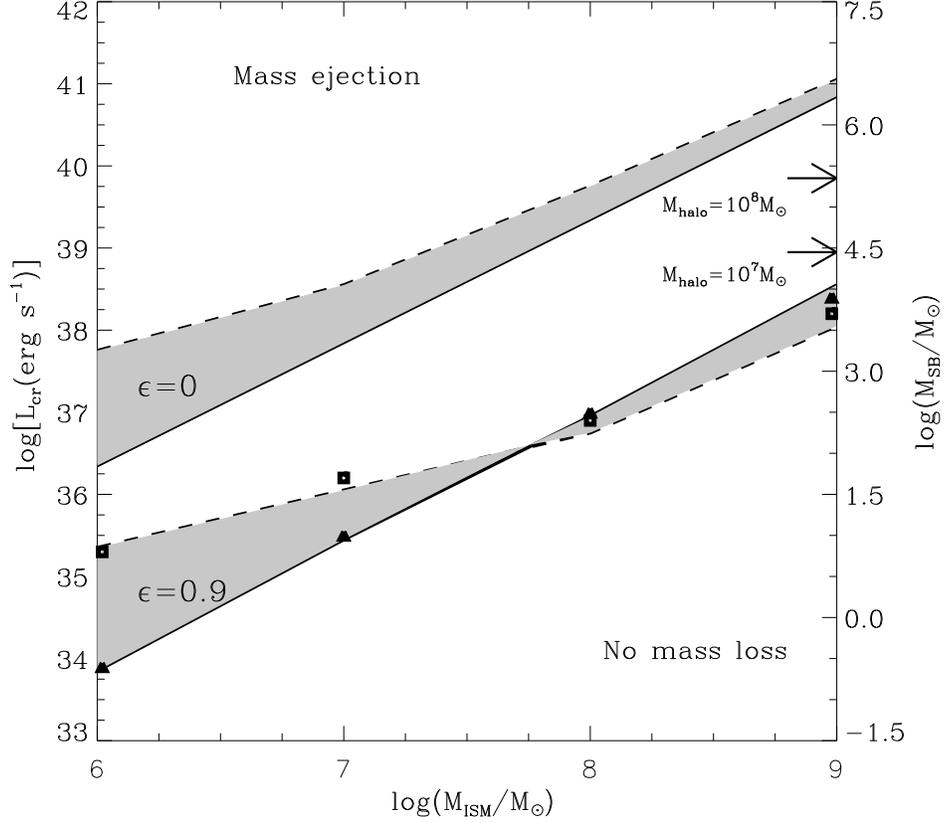}
\caption{Energy estimates. The log of the critical mechanical
luminosity, and of the starburst mass (right-hand axis), required to eject matter from 
galaxies
with a $M_{ISM}$ in the range 10$^6$ -- 10$^9$ M$_\odot$. The lower limit
estimates are shown for galaxies with extreme values of rotation
(typical of spirals) and for two values of the intergalactic pressure  
$P_{IGM}/k$ = 1 cm$^{-3}$ K (solid lines) and $P_{IGM}/k$ = 100 cm$^{-3}$ K
(dashed lines). The upper limit estimates are for galaxies without rotation 
and thus presenting an almost spherical mass distribution. The resolution of our numerical search is
$\Delta log L_{cr}$ = 0.1.
Each line should be considered separately as they divide the
parameter space into two distinct regions: a region of no mass loss 
(below the line) and a region in which blowout and mass ejection
occurs (above the line). Also indicated on  the
right-hand axis are the energy input rates required for a remnant to
reach the outer boundary of a halo with mass 10$^8$ M$_\odot$ and one with
mass 10$^7$ M$_\odot$, for the case of a gravitational potential
provided by $M_{DM}$ = 9.1$\times 10^9$ M$_\odot$ that can hold an
$M_{ISM}$ =  10$^9$ M$_\odot$.
The filled squares and triangles represent analytical energy input rate
estimates.}
\label{fig3}
\end{figure}
%---------------------------------------------------------------

%\newpage

Given their large UV photon output and mechanical energy input rate, 
SSCs are now believed to be the most powerful negative feedback agents 
in starburst galaxies, leading not only to a large-scale  structuring of the ISM and to limit star formation, but also to 
be the agents capable of establishing as in M82 a supergalactic wind,  thereby removing processed material 
from galaxies and causing the contamination of the IGM (Tenorio-Tagle et al. 2003).

\subsection{The inner structure of M82}

The biconical outflow of M82, the nearest example of a supergalactic wind (SGW), displays a collection of 
kpc long optical filaments embedded into an even more extended  pool of soft X-ray emission.
The outflow is known to extend even further, reaching the ``H$_\alpha$ cap" at 11 kpc from the nucleus of M82 
(see Devine \& Bally 1999).
Both features have been partly explained either with the results of 
Chevalier \& Clegg (1985) model of an adiabatic, freely expanding, stationary wind
and/or by the remnant of a large-scale  superbubble evolving into the ISM and the halo of the galaxy
(see for example Suchkov et al. 1994).
Both explanations, based on the energetics  of a single stellar cluster, fail to explain the
detailed inner structure of the outflow. 
The elongated filaments for example, 
are now known to emanate from the central starburst and are not the result of a limb brightened superbubble outer structure
(Ohyama et al 2002). Note also that the filaments cannot be reconcile with instabilities 
in the large-scale supershell, driven by matter entreinment,
which in the models occurs at large distances, kpc from the energy source. 
Furthermore, the stationary superwind solution of Chevalier \& Clegg 
leads to a laminar flow and not to gas condensation or to a
filamentary structure at all. Also, as shown by Strickland \& Stevens 
(2000) it has failed to matched the X-ray luminosity of the M82 
superwind. Note also that the adiabatic assumption, central in the 
model of Chevalier \& Clegg, has recently been shown to be inapplicable 
in the case of massive and concentrated starbursts (Silich et
al. 2003, 2004). Another important issue not accounted by most of the 
numerical simulations is the size of the waist of the biconical structure (150 pc radius in the
case of M82), which in all calculated cases under the assumption of a single
source of energy,  (perhaps with the only exception
of Tenorio-Tagle \& Mu\~noz-Tu\~n\'on 1997, 1998, which account for the infall of matter into the central starburst),
also end up with a remnant that presents a wide open waist
along the galaxy plane (see figures in Tomisaka \& Ikeuchi, 1988; Suchkov et al., 1994).  
Further arguments regarding the disagreement between theory and observations
are given in Strickland \& Stevens 2000,  Strickland, Ponman \& Stevens 
1997, and in Tenorio-Tagle et al. 2003.

\subsection{The physics of supergalactic winds}

So far, all calculations in the literature have assumed that the
energy deposition arises from a single central cluster  that spans 
several tens of pc, the typical size of a starburst.
Following however, the indisputable observational findings with HST, we have recently made the first attempt to calculate 
 the hydrodynamics that result from the interaction of the winds from neighboring
young compact clusters present in a galaxy nucleus  
 (see  Tenorio-Tagle et al 2003).
Several aspects were considered in our two dimensional approach  to the problem. 
 Among these, the metallicity
of the superwind matter was shown to have a profound impact on the inner 
structure of supergalactic winds. 
Full three dimensional calculations of the interaction of multiple SSC winds
are now underway.

Several two dimensional calculations using as initial condition CC85 
adiabatic flows have been performed with the explicit Eulerian finite difference code described 
by Tenorio-Tagle \& Mu\~noz-Tu\~n\'on (1997, 1998). This has been
adapted to allow for the continuous injection of multiple winds (see below).

We have considered the winds from several identical SSCs, each with a mechanical energy 
deposition rate equal to 10$^{40}$ erg s$^{-1}$. The energy is dumped 
at every time step within the central 5 pc of each of the sources following the adiabatic solution of Chevalier \& Clegg (1985).
The time dependent calculations  do not consider thermal conductivity but do account for radiative 
cooling, with a cooling law (Raymond et al. 1976) scaled to the  metallicity assumed  for every case. 

Figure 4  presents the results for which the assumed metallicity of
the winds was set equal to  10Z$_\odot$, justified by the high
metallicity outflows expected from massive bursts of star formation (see Figure 1).
The winds from the SSCs are  exposed to suffer multiple interactions with 
neighboring winds and are also exposed to 
radiative cooling. For the former, the issue is the separation 
between neighboring sources and for the latter the local values of 
density, temperature and metallicity. Radiative cooling would
preferably impact the more powerful and more compact sources, 
leading to cold (T $\sim$ 10$^4$ K) highly supersonic streams.

Figure 4 shows  three 
equally powerful ($L_{SC}$ = 10$^{40}$ erg s$^{-1}$) superstellar clusters
sitting at 0, 60 and 90 pc from the symmetry axis. All of them with an 
$R_{SC}$ = 5 pc, produce almost immediately a stream  with a 
terminal velocity equal to 1000 km s$^{-1}$.  At t = 0 yr the 
three clusters are embedded in a uniform low density 
($\rho$ = 10$^{-26}$ g cm$^{-3}$) medium. Thus our calculations do 
not address the development of a superbubble, nor the phenomenon of 
breakout from a galaxy disk or the halo, into the IGM. The initial condition 
assumes that prior events have evacuated the region surrounding the 
superstellar clusters, and we have centered our attention on the
interaction of the supersonic outflows.

Figure 4 shows the development of a SGW  until it
reaches dimensions of one kpc, together with the final temperature 
structure splitted into the four temperature regimes: The regime of H recombination
10$^4$ K - 10$^5$ K, followed by two regimes of soft X-ray emission 
10$^5$ K - 10$^6$ K, and 10$^6$ K - 10$^7$ K and the hard X-ray 
emitting gas with temperatures between 10$^7$ K - 10$^8$ K.

%\newpage

%---------------------------------------------------------------
\begin{figure}[htbp]
%\plotone{fig4.ps}
\vspace{17.0cm}
\includegraphics{gtt4.ps}
\caption{Two dimensional superwinds. The first four panels represent  
cross-sectional cuts along the computational grid showing: 
isodensity contours with a separation $\Delta$ log $\rho$ = 0.1 and 
the velocity field for which the longest arrow represents 10$^3$ km s$^{-1}$.
The following four panels display isotemperature contours, within the 
range 10$^4$ K - 10$^5$ K, 10$^5$ K - 10$^6$ K,
10$^6$ K - 10$^7$ K and 10$^7$ K - 10$^8$ K, respectively.
Each of the superwinds has a  power of $10^{41}$ erg s$^{-1}$ and 
a radius of 5 pc. The evolution of each wind starts from the adiabatic
solution of CC85.  The 
plots displays the whole computational grid: 100 pc $\times$ 1 kpc.
The assumed metallicities were 
Z = 10 Z$_\odot$. The evolutionary times of 
the first four panels is 
1.79 $\times$ 10$^5$ yr, 4.82 $\times$ 10$^5$ yr, 1.05 $\times$ 10$^6$ yr
and 1.39 $\times$ 10$^6$ yr, respectively.}
\label{fig4}
\end{figure}
%---------------------------------------------------------------

%\newpage

The crowding of the isocontours in the figures indicates steep gradient
both in density or in temperature and velocity, and thus traces the 
presence of shocks and of rapid cooling zones. 
 
The interaction of neighboring supersonic winds causes the immediate 
formation of their respective reverse shocks,
and of a high pressure region right behind them. The pressure (and 
temperature) reaches its largest values 
at the base of the interaction plane, exactly where the reverse 
shocks are perpendicular to the incoming streams.
The high pressure gas then streams into lower pressure regions, 
defining together with radiative cooling, how 
broad or narrow the high pressure zones, behind the reverse shocks, are going to be.

This also happens if cooling is fast enough,  the oblique reverse shocks
rapidly acquire a standing location, however in these cases, 
the loss of temperature behind the shocks is compensated by gas
condensation, leading
to narrow, dense and cold filaments. The drastic drop in temperature 
 occurs near the base of the outflow, where  
the gas density is large and radiative cooling is exacerbated. The dense structures are then  launched at considerable 
speeds ($\sim$ several hundreds of km s$^{-1}$) from zones near the plane
of the galaxy.
These dense and cold structures are easy target to the UV radiation produced 
by the superstellar clusters and thus upon cooling and recombination are likely to become photoionized.
Note however that as the free winds continue to strike upon these structures, 
even at large distances from their origin, the resultant cold filaments
give the appearance of being enveloped by soft X-ray emitting streams.  

All of these shocks are largely oblique to the incoming streams and 
thus lead to two major effects: a) partial thermalization and 
b) collimation of the outflow. These effects result from the fact 
that only the component of the original isotropic outflow velocity perpendicular 
to the shocks is thermalized, while the  parallel component 
is fully transmitted and thus causes the deflection of the outflow 
towards the shocks. This leads both, to an efficient  collimation
of the outflow in a general direction perpendicular to 
the plane of the galaxy, and to a substantial soft X-ray emission 
associated with the dense filamentary structure, extending up to 
large distances (kpc) from the plane of the galaxy.
In the figures one can clearly appreciate that the oblique shocks, confronting the originally diverging flows,
lead to distinct regions where the gas acquires very different temperatures, 
allowing for  radiation in different energy bands.

From our results it is clear that a plethora of structure, both in 
X-rays and in the optical line regime, as in M82, may originate from the 
hydrodynamical interaction  of neighboring winds. The interaction leads 
to multiple standing oblique (reverse) shocks and crossing shocks 
able to collimate the outflow away from the plane of the galaxy. 
In our two dimensional simulations, these are surfaces that 
become oblique to the diverging streams and thus evolve into oblique 
shocks that thermalize only partly the kinetic energy
of the winds causing a substantial soft
X-ray emission at large distances away from the galaxy plane. 
Surfaces that at the same time act as collimators, redirecting the winds 
in a direction perpendicular to the plane occupied by the collection of SSCs. 
Radiative cooling behind the oblique shocks 
leads, as soon as it sets in, to condensation of the shocked gas, 
and thus to the natural development of a network of filaments that 
forms near the base of the outflow, and streams  away from the plane 
of the galaxy to reach  kpc scales. Under many circumstances the 
filaments develop right at the base of the outflow and for all cases the 
prediction is that they are highly metallic. Hydrodynamic 
instabilities play also a major role on the filamentary structure.
Nonlinear thin shell instabilities as studied by Vishniac (1994) 
as well as Kelvin Helmholtz instabilities, broaden, twist and generally 
shape the filaments as these stream upwards and reach kpc scales. 

We thus postulate that if a  collection of SSCs is  sitting in a preferential plane, 
 most of the injected energy would be  channeled in a direction  perpendicular to the plane of the host galaxy. 
This is achieved naturally as a consequence of the plethora of oblique and crossing
shocks that redirect the initially isotropic winds. Collimation thus occurs 
without the need of a thick interstellar matter disk or a torus.

Our considerations point at   a 
new set of possible parameters that profoundly impact the development of supergalactic winds. These are:

\begin{itemize}

\item The number and location of superstellar clusters within a galaxy nucleus.

\item The intensity of star
formation, or stellar mass in every superstellar cluster, which defines their mechanical 
luminosity, 

\item The age of the SSC, which impacts on the metallicity and thus on the cooling of the 
ejected matter.   

\end{itemize}

All of these are
relevant new parameters that may promote, as in M82, 
the inner structure of a supergalactic wind: the co-existense of X-rays and dense filaments, even at large
distances from the sources of energy. Parameters that may promote self-collimation and with it the narrow waist of the biconical 
outflow. All of these features have been confirmed with full 3-D calculations, subject of a forthcoming communication.
Note also that our results led to the full analysis of the HST data of M82 and to  
the discovery of 197
stellar clusters in the nucleus of M82 (see Melo et al. 2005).

\section{Positive feedback from massive and compact SSC}

The location of the threshold line  in the $\dot E_{SC}$ {\it vs} size 
($R_{SC}$) diagram (see Figure 5), the line that defines whether or
not a wind is inhibited,  depends on several variables. It depends on the size of 
the star-forming region ($R_{SC}$) and the 
metallicity of the ejected gas, which has a strong impact on the
cooling curve. It also depends on the assumed
$\dot E_{SC}/\dot M_{SC}$ or adiabatic 
terminal speed ($v_{\infty}$) of the wind. 
 The latter is also bound to the usual assumption that the energy 
deposited by SN is always $10^{51}$ erg but the mass of the stars exploding within the cluster ranges  
from, say, 100 M$_\odot$ to
8 M$_\odot$ and so the injection speed (similar to $v_{\infty}$) and the deposited amount of matter are 
also functions of time. Another factor that strongly affects the location of the threshold line 
is the thermalization efficiency ($\epsilon$) which defines the fraction of the mechanical energy that
upon thermalization, 
can be evenly spread within the cluster volume. Estimates of $\epsilon$ by several 
authors lead to values between 1 (Chevalier \& Clegg; 1985) and 
0.03 (see Melioli \& Del Pino 2004 and references therein) and
depends simply on the proximity of the sources undergoing winds and SNe, which through radiation may reduce the amount of 
energy available after thermalization.   
We have shown that there are three different types of solutions: SSCs far away from the threshold line (low mass,
low energy clusters) 
undergo a quasi-adiabatic evolution well described by the Chevalier \& Clegg (1985) solution. More energetic clusters 
are to have strongly radiative winds. Cooling hardly affects their velocity ($v_w \sim v_{\infty}$) 
and density distribution ($\rho_w \propto r^{-2}$), but
their temperature instead of falling as $r^{-4/3}$, it 
falls rapidly to $T_w \sim 10^4$ K close to the SC boundary and the more so, 
the closer they are to the threshold line (see Figure 2). The strongly radiative winds  around such clusters
lead, compared to the adiabatic solution,  
to very much reduced X-ray envelope sizes. The third solution is for clusters  above the threshold line.
These would have their 
winds inhibited and as shown below, this  turns them into very efficient positive feedback star-forming agents.

%\newpage

\begin{figure}[htbp]
\plotone{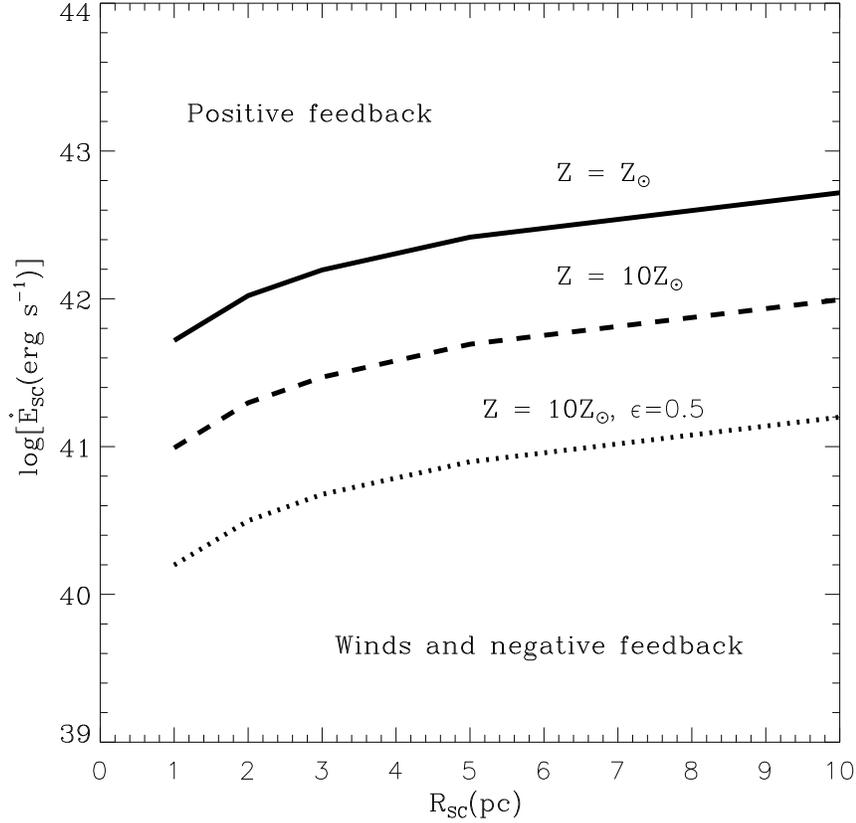}
\caption{The threshold line. The location of the threshold line, the
line that divides the $\dot E_{SC}$ {\it vs} size ($R_{SC}$)
diagram into two distinct areas. There the matter deposited within the
SSC volume, through winds and SNE, ends up either streaming away as a 
stationary (adiabatic or radiative) wind (below the line) or it
accumulates to end up steadily being driven into new episodes of star formation
(above the line). Different locations of the threshold line, for an
assumed full thermalization efficiency ($\epsilon$ = 1),  are display
for an ejecta metallicity value equal to solar and ten times solar. 
The change in location  for different values of $\epsilon$ is also indicated.}
\label{fig5}
\end{figure}
%---------------------------------------------------------------

%\newpage

The facts above the threshold line are that radiative cooling drastically diminishes 
the sound speed $c_{SC}$ and the pressure gradient across the SSC volume, 
inhibiting the possibility of a wind. Radiative cooling upsets then the  
stationary condition in which the deposited matter ($\dot M_{SC}$) has to equal, at all times, the amount 
of matter streaming out of the SC volume

\begin{equation}
\dot M_{SC} = 2 \dot E_{SC}/v_\infty^2 = 4 \pi R_{SC}^2 \rho_{SC} c_{SC}
\end{equation}

\noindent where $v_\infty$ 
is the resultant wind terminal speed in the absence of radiative cooling. 
As soon as  this happens, the mass returned by the stars 
($\dot M_{SC}$) begins to accumulate, promoting 
larger densities and an even faster cooling within the SSC volume. Following this trend,  the accumulated gas density ultimately 
fulfill the gravitational instability criterion causing the collapse into a new stellar generation.

For clusters above the threshold line, 
Figure 6 shows how the density of the accumulating gas
($\rho_{ac} = 3 \dot M_{SC} t/ 4 \pi R_{SC}^3$; where  $t$ 
the evolution time), grows as a function of time  within the SSC volume, 
until the moment when the density of the accumulating gas  
exceeds the gravitational instability criteria 
%--------------------------------------------------------------- 
\begin{equation}
\label{jeans}
\rho_J \sim 2.3 
\times 10^{-20} \left(\frac{T}{100K}\right) \left(R_{SC1}\right)^{-2} \; g \, cm^{-3}
\end{equation}
%--------------------------------------------------------------- 
where $\rho_J$ is the Jeans density and  $R_{SC1}$ is the SSC radius 
in pc units. At that moment, when $\rho_{ac} = \rho_J$ collapse will inevitably proceed. Note that for a given SSC 
(with a fixed volume), $\rho_{Jeans}$
is only a function of temperature and this is set, at least initially, by photoionization.

%\newpage

\begin{figure}[htbp]
\plotone{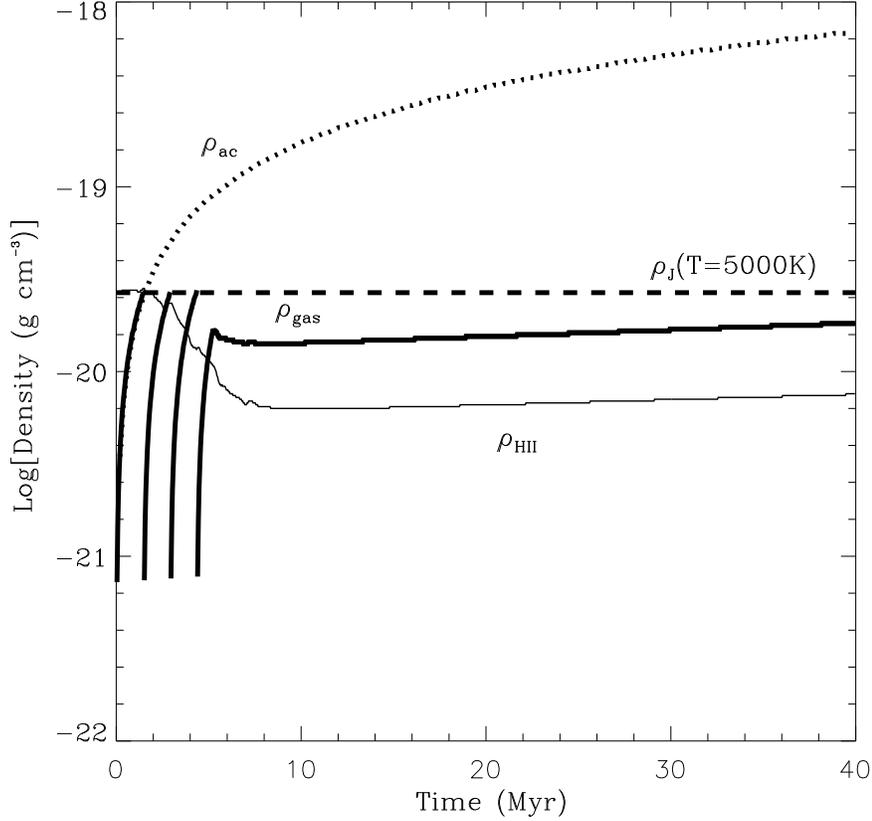}
\caption{Positive star-forming feedback. The figure 
shows the rapid growth of the reinserted gas density within the SSC volume
(rising solid lines), interrupted as it reaches the gravitational
instability criterion ($\rho_J$), leading to a new stellar generation
and to a new phase of matter accumulation.  The situation changes
slightly as the available photon flux becomes unable to
sustain all the SSC volume fully ionized, allowing for recombination
and further cooling of the recombined matter, and thus to a much lower
limit of the Jeans instability criterion. This allows the accumulation
time to become comparable to the free-fall time, establishing a new
stationary condition in which $\dot M_{SC}$ = SFR. The density value 
($\rho_{gas}$) required for this condition is also indicated in the 
figure as well as the maximum density value ($\rho_{HII}$)
that can be supported fully ionized within the SSC volume, with the 
available UV photon flux.}
\label{fig6}
\end{figure}
%---------------------------------------------------------------

%\newpage

For massive clusters the initial ample supply of 
UV photons exceeds at first the number of recombinations within the volume occupied 
by the reinserted gas and the resultant HII region,  given the large metallicities of the ejecta, 
is here assumed to rapidly approach an  
equilibrium temperature 
$T_{HII} \sim$ a few $ 10^3$ K. At these temperatures the sound
speed ($<$ 10 km s$^{-1}$)
remains well below the escape speed  and the
reinserted gas would inevitably continue to accumulate to rapidly (within $1.5 \times  10^6$ yr) 
reach the value of the Jeans density 
for a gas at say, $T_{HII}$ = 5000 K, and collapse into a new stellar generation. The event gives 
rise to a new phase of matter accumulation, which once more will rapidly approach $\rho_J$ (for $T$= 5000 K) and undergo collapse 
within a free-fall time, of the order of 10$^5$ yr, while transforming  $\sim 2 \times 10^5$ M$_\odot$
into stars. All stellar generations resultant from mass accumulation 
within the SSC volume
have here been assumed to also acquire a Salpeter IMF with similar upper and lower 
mass limits as those imposed to the main superstellar cluster,
and their resultant properties (mechanical energy and UV photon output) have been added to those produced by the main cluster.

A few, almost identical, stellar generations 
 are expected  from the accumulation process (solid rising lines in Figure 6), every time that the accumulated gas density
 $\rho_{ac}$ reaches $\rho_J(5000K)$ (dashed line in Figure 6). 
 The situation changes slightly when the number of ionizing photons ($N^0$), 
despite the added contribution  
of secondary stellar generations,  
 becomes insufficient to fully ionize the matter accumulated  within the star cluster volume. This is due to the 
evolution of the main cluster, whose UV photon output begins to fall as $t^{-5}$ after $\sim$ 3.5 Myr. 
Figure 6 shows $\rho_{HII}$ (thin solid line), the maximum  density within  the SSC volume that  can be 
supported fully ionized by the UV radiation produced by the evolving cluster
($\rho_{HII} = (3 N^0 \mu^2/(4 \pi R_{SC}^3 \beta))^{0.5}$; where $\mu = 1.4 m_H$ and $\beta$, the recombination coefficient
to to all levels above the ground level = 2.59 $\times 10^{-13}$ cm$^{-3}$ s$^{-1}$).
During the accumulation process, once $\rho_{ac}$ exceeds $\rho_{HII}$, the ionized volume begins to shrink
to end up as a collection  of
ultra compact HII regions around the most massive stars left within
the cluster, while the bulk of the ejected material, now recombined,  continues to cool, 
approaching rapidly a temperature $\sim$ 100 K. 
The characteristic cooling time 
($t_\Lambda = 3kT/2 \Lambda n$; where $\Lambda$ is the cooling rate, 
a function of T and Z) is also very small, $\sim$ 
1.5 $\times 10^5$ yr, and is to become even shorter as matter 
continues to accumulate. 
Matter is at all times uniformly replenished within the whole SSC volume, and thus the gas density
presents an almost  uniform value. 
However,  the accumulating gas  now has two different temperatures ($T_{HII}$ and 100 K) and as $\rho_{ac}$ grows and 
the fraction of the ionized volume ($f_{HII} = (3N^0\mu^2)/(4\pi R_{SC}^3 \beta \rho_{ac}^2$)) shrinks, the size of  
cold  condensations (at 100 K) able to  become gravitationally unstable and their free-fall time also become smaller.

The drop in the number of ionizing photons  and the consequent growth of the 
neutral volume,  lead then to a  second important condition
in which the  
characteristic accumulation time 
%--------------------------------------------------------------- 
\begin{equation}
\label{eq.1} 
\tau_{ac} = \frac{4}{3} \pi \rho_{gas}   R_{SC}^3 (1 - f_{HII})/ {\dot M} 
\end{equation}  
%---------------------------------------------------------------
becomes equal to the free-fall time 

%--------------------------------------------------------------- 
\begin{equation}
\label{eq.2}
t_{ff}= \sqrt{\frac{3 \pi}{32 G \rho_{gas}}} ,
\end{equation}
%--------------------------------------------------------------- 
This condition 
defines $\rho_{gas}$, the density  above the Jeans instability limit for a neutral condensation at 100 K  ($\rho_J$(100K)): 

%--------------------------------------------------------------- 
\begin{equation}
\label{eq.3} 
\rho_{gas} = \left(\frac{27 {\dot M}^2}{512 \pi G \left(1 -f_{HII} \right)} \right)^{1/3} 
             R_{SC}^{-2} = 1.15 \times 10^{-18} \left(\frac{\dot M}{1M_\odot yr^{-1}} 
             \right)^{2/3} \left(\frac{R_{SC}}{1pc}\right)^{-2} 
\end{equation}  
%--------------------------------------------------------------- 
and thus once $\rho_{ac}$ becomes equal to $\rho_{gas}$, a new stationary solution becomes possible.

Everything happens  very rapidly, compared to the evolution time-scale  
of the parental cluster ($\sim$ 40 Myr), and almost at the same time. The ejected matter is 
thermalized within the SSC volume and immediately begins to cool. At  the same time that it 
accumulates making cooling even faster. This allows it to rapidly reach the required $\rho_{gas}$ value, above the
Jeans instability limit, that warrants its collapse in a similar time-scale, 
while the collapsing material is replenished by the newly ejected matter.
When this happens, the mass deposition rate from the cluster becomes equal to
the rate of star formation. 
Gravitational collapse and star formation within the star cluster volume and with the
matter injected by all sources, drive in this way a new stationary condition through  a new era of quasi-continuous star formation
in which $\dot M_{SC}$ is now equal  to the star formation rate (SFR). Further details 
describing this phase are given in Tenorio-Tagle et al. 2005.

\section{Conclusions}

Superstellar clusters are certainly the main mode of massive star formation in starburst and
interacting galaxies. We have reviewed here how is that they work and the possible impact that they may have into the 
surrounding ISM. We have revised the assumptions of Chevalier \& Clegg (1985) dealing with thermalization 
and the  flow requirements to establish a stationary superwind emanating from these sources. 
We have also emphasized the importance of radiative cooling and how does it affect the superwind X-ray envelopes.

Calculations in the literature have left  clear the fact that single energy sources lead to superbubbles and supershells.
However, to produce a supergalactic wind with a detailed inner structure as in M82,  
multiple sources, seem to be required. 

We have also shown that a straight forward extrapolation towards the most massive and compact coeval clusters
is not valid. When radiative cooling becomes significant within the SSC volume itself, then 
 instead of 
driving a superwind able to disperse the surrounding ISM
and even channel  its way into the IGM, events that have make them been regarded as negative feedback agents, they 
become  in fact extreme examples of positive star formation feedback.

Massive and compact coeval clusters appear in the $\dot E_{SC}$ {\it vs} cluster size diagram above the threshold line,
in the region where radiative cooling inhibits the development of stationary superwinds. 
In such cases the matter reinserted,  through stellar winds and supernovae, is unable to escape
and after a short phase of
matter accumulation,  a new stationary solution in which $\dot M_{SC}$ becomes equal to the SFR is rapidly met. 
A positive feedback condition in which 
new stellar generations result in situ, from the collapse of the
matter reinserted  by the sources evolving within the star cluster volume.

The massive concentrations
imply a high efficiency of star formation which permits even after long evolutionary times  the  
tight configuration that characterizes them, despite stellar evolution and its 
impact through photo-ionization, winds and supernova explosions,
believed to efficiently disperse the gas left over from star formation.
It is thus the self-gravity that results from the high efficiency what keeps the sources bound together.

The secondary star formation process while causing a faster mass deposition rate, drives the SFR to grow 
from 0.1 to 0.25 M$_\odot$ yr$^{-1}$ over the parent cluster supernova phase ($\sim $ 40 Myr).
The continuous reprocessing of the ejected
material leads effectively to a continuous transformation of the high mass
stars into a low mass ($\leq$ 8 M$_\odot$) population,  keeping constant the total mass of the stellar
component.

A central issue, regarding ISM studies, is the fact that the more massive and compact
clusters (as those detected by HST and other large ground-based telescopes), are unable to generate superwinds and  
shed their metals into the ISM or the IGM.
Their evolution leads to many stellar generations and thus to a mixture of stellar populations, 
all contaminated by the products from former stellar generations.
An exacerbated episode of star formation that leaves no trace of its evolution in the ISM.

\acknowledgments

GTT acknowledges finantial support from the Secretar\'\i{}a de Estado de Universidades e Investigaci\'on (Espa\~na)
ref: SAB2004-0189 and the hospitality of the Instituto de Astrof\'\i{}sica de Andaluc\'\i{}a (IAA, CSIC) in Granada, Spain.
This study has been partly supported by 
AYA2004-08260-CO3-O1 from the Spanish Consejo Superior de Investigaciones Cient\'\i{}ficas.

%\newpage

%\onecolumn


%\begin{figure}
%\figcaption[ms15457_1.ps]
%{The dispersion relation $\omega (\eta )$ for $c_{sh}$ corresponding to 
%position of a 100 M$_{\odot}$ fragment. \label{fig1}}
%\end{figure}


\begin{thebibliography}{99}

\bibitem{1} Cervi\~no, M. \& Mas-Hesse, M. 1994, A\&A 284, 749
\bibitem{2} Chevalier, R.A. \& Clegg, A.W. 1985, Nature, 317, 44
\bibitem{3} Colina, L., Gonzalez-Delgado, R., Mas-Hesse, M. \& 
            Leitherer, C. 2002 ApJ 579, 545 
\bibitem{4} Devine, . \& Bally, J. 1999 ApJ 510, 197
\bibitem{5} Ho, L. C. 1997, Rev.MexAA, Conf. Ser. 6, 5
\bibitem{6} Johnson, K. E., Kobulnicky, H. A., Massy, P.  \&  
            Conti, P. S. 2001, ApJ 559, 864
\bibitem{7} Kobulnicky, H. A. \& Johnson, K. E. 1999, ApJ 527, 154            
\bibitem{8a} Larsen, S. S. \& Richtler, T. 2000, A\&A 354, 836 
\bibitem{8b} Lamers, H., et al., 2004, in ASP Conf. Ser. 322, The formation and 
      evolution of massive young star clusters,  
      eds. H.J.G.L.M. Lamers, L.J. Smith \& A. Nota
      (San Francisco: ASP)
\bibitem{9} Leitherer, C., Schaerer, D., Goldader, J.D. et al. 1999,
            ApJS, 123, 3
\bibitem{9a} Melioli, C. \& de Gouveia Del pino E. M. 2004 A\&A 424, 817
\bibitem{9b} Melo, V., Mu\~noz-Tu\~n\'on, C., Maiz-Apellaniz, J. \&  
             Tenorio-Tagle G. 2005 ApJ 619, 270
\bibitem{10} Meynet, G. \& Maeder, A. 2002, A\&A, 390, 561.
\bibitem{11} Ohyama, Y. et al. 2002, in the Proceedings of the IAU 8th 
             Asian Pacific Regional Meeting Vol 2. Eds. S. Ikeuchi, 
             J. Hearnshaw, T. Hanawa (Tokio ASJ), 285.
\bibitem{12} Pasquali, A., Gallagher, J. S. \& de Grijs, R. 2004 A\&A 415, 103
\bibitem{13} Raymond, J. C., Cox, D. P. \& Smith, B. W. 1974 ApJ 204, 290.
\bibitem{14} Silich, S. \& Tenorio-Tagle G. 2001 ApJ 552, 91.
\bibitem{15} Silich, S. \&  Tenorio-Tagle G. 
             Mu\~noz-Tu\~n\'on, C. 2003, ApJ, 590, 796.
\bibitem{16} Silich, S., Tenorio-Tagle G. \& 
             Rodr\'\i{}guez Gonz\'alez, A. 2004, ApJ, 610, 226.
\bibitem{17} Silich, S., Tenorio-Tagle G., Terlevich, R., Terlevich, E. \& 
             Netzer, H. 2001, MNRAS, 324, 191
\bibitem{18} Strickland, D.K., Ponman, T. J. \& Stevens, I. R., 1997, 
             A\&A,  320, 378             
\bibitem{19} Strickland, D.K. \& Stevens, I. R., 2000, MNRAS, 314, 511
\bibitem{20} Suchkov, A. A., Balsara, D. S., Heckman, T. M. Leitherer, C. 
             1994, ApJ, 463, 528
\bibitem{21} Tenorio-Tagle, G. \& Mu\~noz-Tu\~n\'on, C. 
             1997, ApJ, 478,  134
\bibitem{22} Tenorio-Tagle, G. \& Mu\~noz-Tu\~n\'on, C. 
             1998, MNRAS, 293, 299 
\bibitem{23} Tenorio-Tagle, G., Silich, S. \& Mu\~noz-Tu\~n\'on, C. 
             2003, ApJ, 597, 279
\bibitem{24} Tenorio-Tagle, G., Silich, S., Rodr\'iguez-Gonz\'alez A. 
             \& Mu\~noz-Tu\~n\'on,  C., 2005, ApJ 620, 217.
%\bibitem{17} Tenorio-Tagle, G., Silich, S. Rodr\'\i{}guez Gonz\'alez, A. 
%             \& Mu\~noz-Tu\~n\'on, C. 2005, ApJL, (in press)
\bibitem{25} Tomisaka, K. \& Ikeuchi, S. 1988 ApJ, 330, 695.
\bibitem{26} Walcher, C.J., van der Marel, R.P., McLaughlin, D., Rix,
             H.-W., B\"oker, T., H\"aring, N., Ho, L.C., Sarzi, M. \&
             Shields, J.C. 2005, ApJ, 618, 237
\bibitem{27} Whitmore, B. C., Zhang, Q., Leitherer, C., Fall, M. S., 
             Schweizer, F. \& Miller, B. W., 1999, AJ, 118, 1551             

\end{thebibliography}
\end{document}